\documentclass[epsfig,12pt]{article}
\usepackage{graphicx}
\usepackage{hyperref}
\begin{document}
\begin{center}
{\large\bf  Radiative Relaxation Time of Longitudinal Polarization in the $2\times 3\times  5.7 ^0$ Central Arc Sheme for SuperB\footnote{This work was done in 2008 at INFN (Frascati) for the SuperB project and was registered as internal report SB-NOTE-ACC-2008-001.}}
\end{center}
\begin{center}
{S.A. Nikitin\\ BINP, Novosibirsk, RF}
\end{center}
\section{Introduction}
We consider one of the variants of obtaining the longitunal polarization of electrons in the SuperB HER at 7 GeV. 
Spin vector manipulation in this scheme is carried out with the help of two spin rotator insertions on the left and on the right ends of the FF region
and $2\times 5.7\times 3 ^0$  bend magnets ("Central Arc") arranged symmetrically between the insertions relative to I.P..
Each insertion contains 4 similar solenoids rotating a spin in common by $\pi/2$ and five quadrupole lenses for betatron coupling
localization (proposed by U. Wienands \cite{panpri}). In the paper the neccesary spin-orbit coupling coefficients are found. The time of polarization relaxation
due to SR processes in the bend magnets of main arcs is calculated depending on the beam energy region of 7 GeV. 
Comparison between
the radiative relaxation time of polarization and the beam lifetime due to high luminosity enables to estimate a convenience of the longitudinal
polarization scheme. The scheme is compared in short with one of the alterntive variants \cite{bgmnik}.

\section{Spin kinematics}

\begin{figure}[htb]
\centering
\includegraphics*[width=100mm]{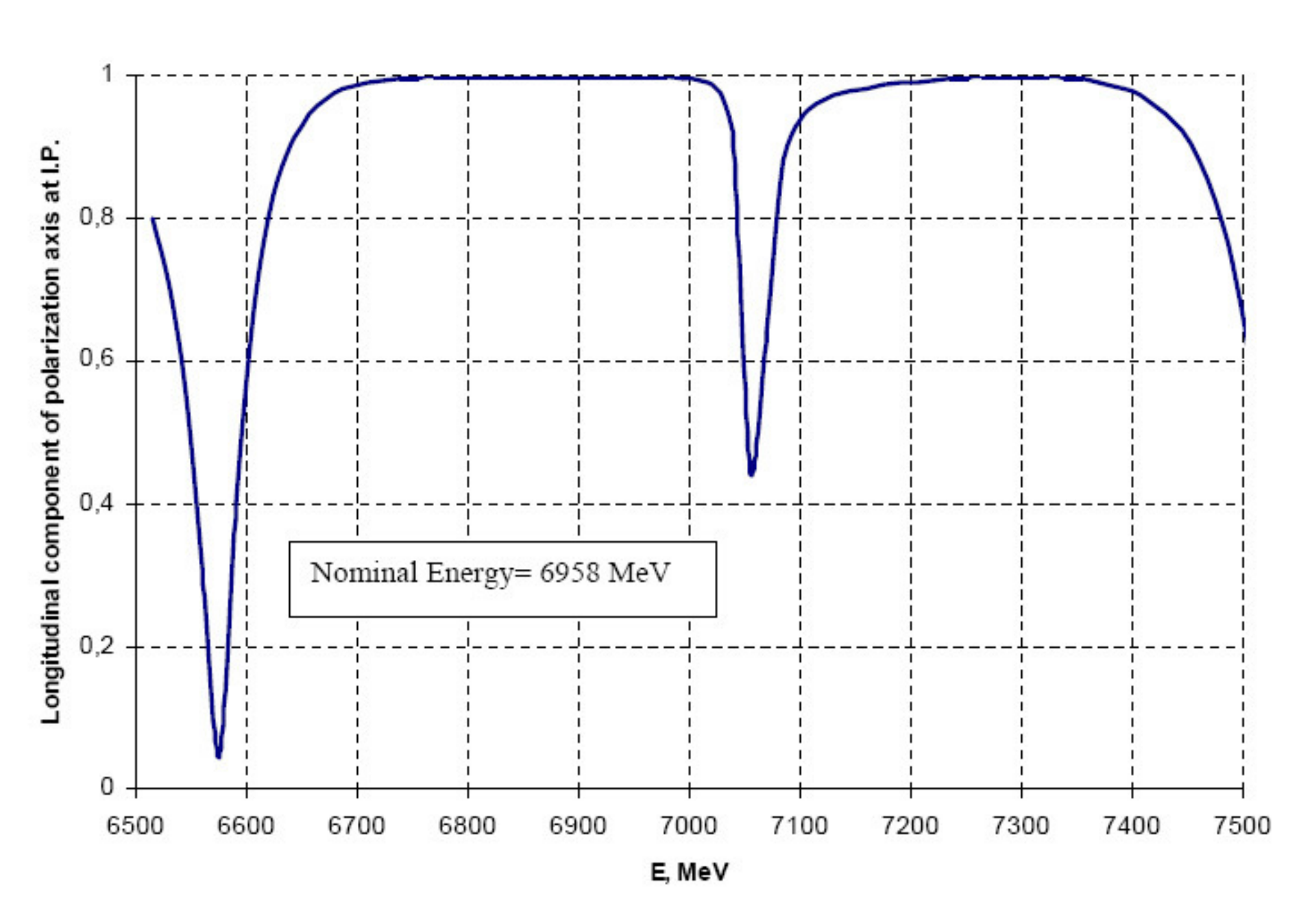}
\caption{\it Energy dependence of the longitudinal projection of polarization axis at I.P.. Central Arc bend is $2\times 3\times 5.7^0$.}

\label{f1}
\end{figure}

The kinematic scheme provides a dynamically stable longitudinal direction  of the polarization axis $\vec n$ at I.P.  and  at the same time
it restores vertical polarization in the main HER arcs. This is exactly valid only at a nominal beam energy value which corresponds to a spin
rotation angle of $(2k+1)\pi/2$ over the half Central Arc bend magnets (assuming a rotation in the single solenoid insertion by $\pi/2$). 
Recently it has been shown for such a scheme \cite{bgmniktau} that the longitudinal  projection
of polarization axis at I.P. remains very close to unity over some region around  of a nominal energy. Furthermore, a similar picture 
- quasi-flat dependence with holes near the energy points where the $\nu=E/440.65$ is close to integer values - is repeated with changing energy.
Energy dependence of the longitudinal component of the $\vec n$ axis at I.P. for the scheme under consideration is plotted using
the spin matrix technique in Fig.~\ref{f1}. Nominal energy is 6958 MeV.

\begin{figure}[htb]
\centering
\includegraphics*[width=100mm]{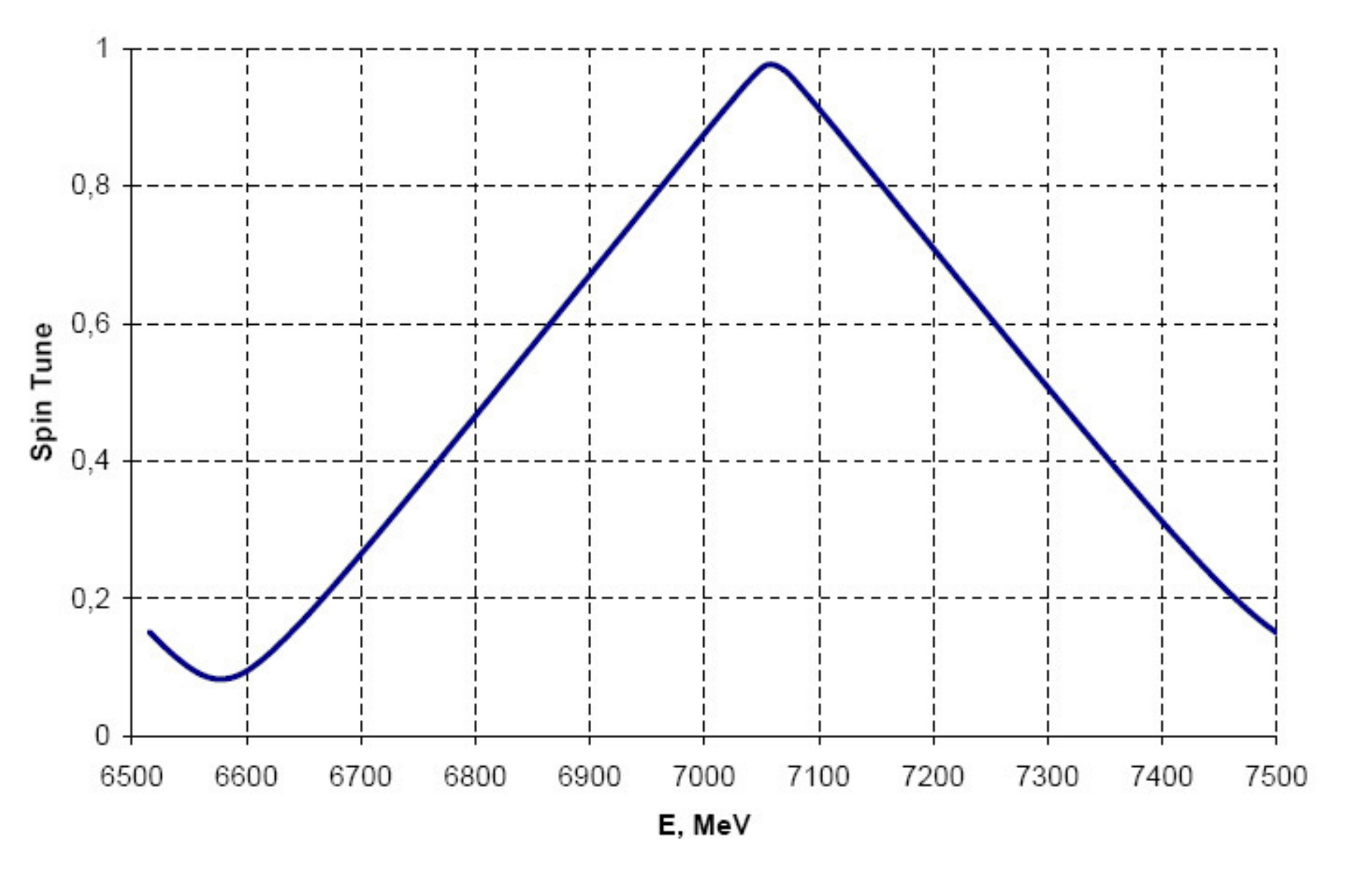}
\caption{\it Spin tune versus the beam energy for the $2\times 3\times 5.7^0$ Central Arc scheme.}

\label{f2}
\end{figure}

Spin tune $\nu_0$ defined as a non-integer part of the spin precession frequency in units of
revolution frequency is approximately a non-integer part of the $\nu$-parameter value in the vicinity of nominal energy. Generally, 
the spin tune differs from that parameter (see Fig.~\ref{f2}). Note, the Central Arc Scheme excludes integer spin resonances for the polarization axis:
the spin tune do not become definitely integer anywhere at energy. It may just approach very close to integer values as seen in Fig.~\ref{f2}.
Closeness of the effective spin tune to integer values as well as to integer combinations with betatron tunes leads to increase of spin-orbit coupling
and, as a sequence,  depolarization effect of synchrotron radiation.

\section{Solenoid insertion optics}
\begin{figure}[htb]
\centering
\includegraphics*[width=120mm]{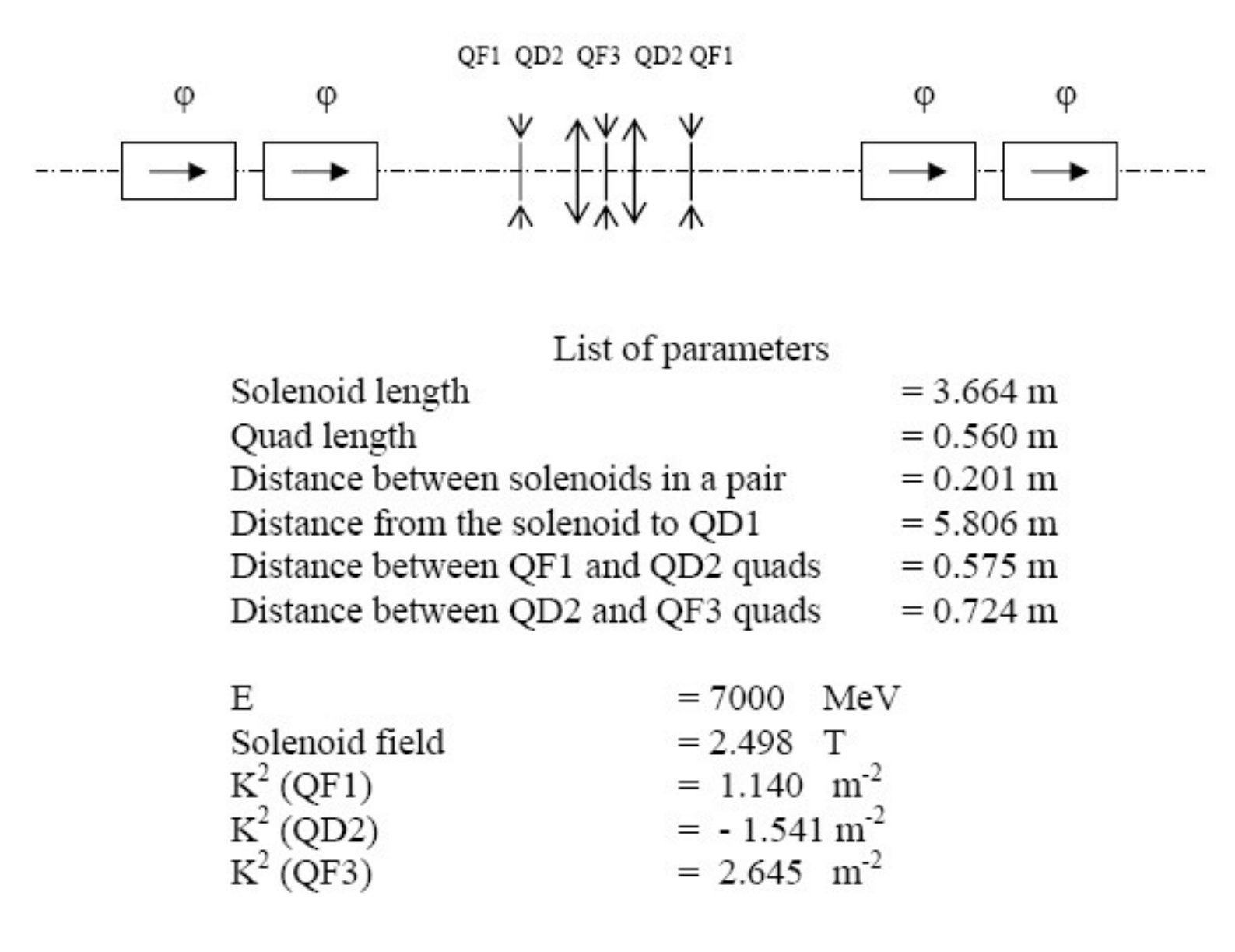}
\caption{\it The solenoid insertion optics scheme and list of its parameters.}

\label{f3}
\end{figure}

Magnetic structure of the solenoid insertion proposed by U. Wienands (see Fig.~\ref{f3}) fulfills the requirement of betatron coupling localization.
Transfer matrix $4\time 4$ for the vector $(x, x', y, y')$ from an input point of the insertion to output one has a form with zero adiagonal blocks:
\[\left(\begin{array}{cccc}
0.053452 & 0 & 0&0 \\
0 & 18.708287&0&0 \\
0 & 0 & 0&-0.118322 \\
0 & 0 &8.451543 &0 \\
\end{array} \right)\]
Dispersion function and its derivative are identically equal zero over the section with insertion. The insertions are incorporated into the general
HER magnetic structure with a minimal influence on chromatic effects \cite{panpri}.   

\section{Polarization kinetics formulas}
\subsection{Radiative polarization}
Generally, an asymptotic (equilibrium) degree of the radiative polarization with the arbitrary vectorial functions of azimuth $\theta$ - the polarization axis
$\vec n(\theta)=\vec n(\theta+2\pi)$ and the spin-orbit coupling $\vec d (\theta)=\vec d (\theta+2\pi)$ is given by the formula \cite{Derb}
$$\zeta=\frac{8}{5\sqrt{3}}\frac{<|\dot{\vec v}|^2\vec v \times \dot{\vec v}(\vec n-\vec d~  )>}{<|\dot{\vec v}|^3[1-\frac{2}{9}(\vec n\cdot\vec v)^2+\frac{11}{18}{\vec d}^2]>}, \eqno{(1)}$$
where $\vec v$ is a velocity vector ($c=1$); $\vec v \times \dot{\vec v}\propto \vec H_\perp$, $\vec H_\perp$ is the transverse magnetic field;
the angular brackets mean averaging over the azimuth. Corresponding relaxation time is
$$\tau_r=\tau_0 \frac{<|\dot{\vec v}|^3>}{<|\dot{\vec v}|^3[1-\frac{2}{9}(\vec n\cdot\vec v)^2+\frac{11}{18}{\vec d}^2]>}, \eqno{(2)}$$
$\tau_0$ is the radiative polarization rise-time for a given storage ring when all special insertions diverting the polarization vector from
 the  vertical axis are turned off. It may be estimated from
$$\tau_0=2.74\times 10^{-2} \frac{\rho^2 R}{E^5},$$
where $\tau_0$ is in hours; $\rho$,  the bend radius of magnets in arcs,  and  $R$, the mean machine radius, are in meteres; 
the beam energy $E$ is in GeV.

\subsection{Steady polarization degree under Trickle Injection}  
It is assumed that the polarized with a degree of $P_0$ electrons are injected  into the SuperB ring from a linac in the so-called Trickle Injection mode.
The frequency $f_i=1/\tau_i$ of injection per a single bunch is sufficiently high 
to compensate the particle loss caused mainly by a high luminosity and characterized by beam lifetime $\tau_{\it l}$ ($\tau_i<<\tau_{\it l}$).
If radiative
relaxation of polarization is depolarizing process ($\tau_r<<\tau_0$) and it dominates over other similar effects, for instance, the beam-beam depolarization
\cite{nikBB},  the polarization degree ($\bar{P}$) averaged over the injection cycle can be found from
$$\bar{P} \approx P_0 \frac{\tau_r}{\tau_r+\tau_{\it l}} \eqno{(3)}$$	
In a case the particle loss rate is notably greater than the radiative depolarization rate ($\tau_l<<\tau_{\it l}$) the average polarization
degree is close to the polarization degree of injected particle: $\bar{P}\approx P_0$.  Generally, one must take into account the asymptotic
radiative polarization degree $\zeta$ (I. Koop, \cite{CDR})
$$\bar{P}=P_0\frac{\tau_r}{\tau_r+\tau_{\it l}}+\zeta\frac{\tau_{\it l}}{\tau_r+\tau_{\it l}}.$$
In the case under consideration, the radiative spin kinetics is determined basically by quantum fluctations in the main arcs where
the polarization is vertical.
By this reason, the relaxation time and eqilibrium degree of polarization are related approximately like they are in the conventional storage rings:
$$\zeta\approx 0.9\frac{\tau_r}{\tau_0}.$$
Therefore,  
$$\bar{P}\approx\frac{\tau_r}{\tau_r+\tau_{\it l}}\left(P_0+0.9\frac{\tau_{\it l}}{\tau_0}\right).$$
For the SuperB Factory the condition $\tau_{\it l}<<\tau_0$ is fulfilled  ($\tau_
{\it l}$ is about a few of minuts and $\tau_0$ is more than 7 hours at 7 GeV). So the steady polarization degree is determined rather 
through the ratio $\tau_r/\tau_{\it l}$ as well as the injected particle polarization $ P_0$ and it does not depend upon the asymptotic radiative
polarization degree $\zeta$.

\section{Spin-orbit coupling coefficients calculation}

\begin{figure}[htb]
\centering
\includegraphics*[width=120mm]{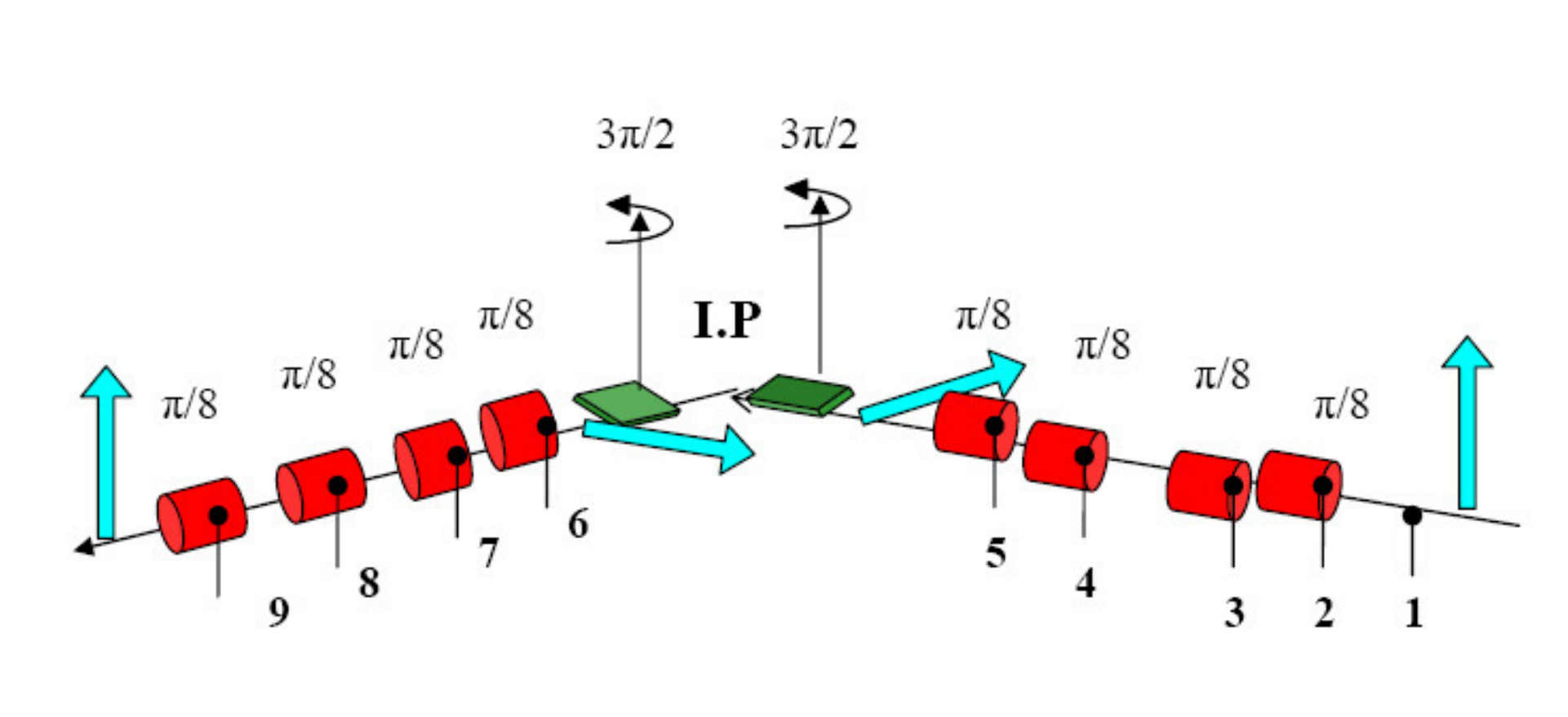}
\caption{\it Reference points in the polarization scheme of 8 solenoids with the $2\times 3\times 5.7 ^0$
 Central Arc bend.
At the points 2, 3, ... 9 the fringing field of the solenoids ends and their uniform field region begins. 
}

\label{f4}
\end{figure}

Bend magnets in the SuperB main arcs make together a main contribution $\propto$ the magnetic field cubed to the radiative spin kinetics according to the equations (1) and (2).
The Central Arc magnets have the field of a similar order with the main arc magnets but take only small segment of the total circuit. By this reason, it is enough to account
as a first approximation the spin orbit coupling in the main arcs only. (Finally, the central bend magnet contribution can be taken into account using formulas obtained in \cite{nik1,nik2}.
The general expressions of the $\vec d(\theta)$ for the scheme in Fig.~\ref{f4} was obtained in \cite{nik1}. In particular, the interesting value is 
$$|\vec d (\theta)|^2=[\Re(T)]^2+[\Im(T)]^2, \eqno{(4)}$$
$$\Re(T)=-\frac{|h|\beta_{x1}^{-1/2}}{2 \sin{\pi(\nu_x+\nu_0)\sin{\pi(\nu_x-\nu_0)}}}\left\{(A\beta_{x1}-B\alpha_{x1})[\cos{\psi}\cos{2\pi\nu_0}-\cos{(2\pi\nu_x+\psi)}]+\right.$$
$$\left. +B[\sin{(2\pi\nu_x+\psi)}-\sin{\psi}\cos{2\pi\nu_0}]\right\}-\frac{1}{2}D,$$
$$\Im(T)=-\frac{|h|\beta_{x1}^{-1/2}\sin{2\pi\nu_0}}{2 \sin{\pi(\nu_x+\nu_0)\sin{\pi(\nu_x-\nu_0)}}}\left\{-(A\beta_{x1}-B\alpha_{x1})\cos{\psi}+B\sin{\psi}\right\}-\frac{1}{2}D\cot{\pi\nu_0}.$$
Here, 
$$h(\theta)=h(\theta+2\pi)=|h|e^{i\psi},~ ~  |h|=\frac{1}{2}\sqrt{{\cal H}},~ ~ {\cal H}(\theta)=\frac{\eta_x^2+(\alpha_x\eta_x+\beta_x\eta_x')^2}{\beta_x},$$
$$\psi(\theta)=\arccos{\left(-\frac{\eta_x}{\sqrt{{\cal H}\beta_x}}\right)-\int\limits_0^\theta \frac{R d\theta'}{\beta_x}},$$
$\alpha_x$, $\beta_x$, $\eta_x$, $\eta_x'$ are the radial amplitude and dispersion functions as well as their derivatives; $\beta_{x1}$, $\alpha_{x1}$ are the values at origin point (see Fig.~\ref{f4});
$\nu_x$ is a radial betatron tune, $\nu_0$ is a spin tune which in the case under consideration coincides with the spin tune value for the conventional storage ring:
$\nu_0=\nu=E[MeV]/440.65$. The coefficients $A,~ B$, and $D$ are have been modified as compared with \cite{nik1,nik2} due to differences in the solenoid-based insertion structure: 
$$\eqno{(5)}$$
\[\left(\begin{array}{c}
A \\ B
\end{array} \right)=\nu\varphi\left[-\left(\begin{array}{c}
s_2-u_6 \\ t_2-v_6
\end{array} \right)+\sin{\varphi}\left(\begin{array}{c}
u_3+s_7 \\ v_3+t_7
\end{array} \right)-\cos{\varphi}\left(\begin{array}{c}
s_3-u_7\\ t_3-v_7
\end{array} \right)+\right. \]
\[\left.+\sin{2\varphi}\left(\begin{array}{c}
u_4 +s_8\\ v_4+t_8
\end{array} \right)+\cos{2\varphi}\left(\begin{array}{c}
-s_4+u_8 \\ -t_4+v_8
\end{array} \right)+sin{3\varphi}\left(\begin{array}{c}
s_9+u_5 \\ t_9+v_5
\end{array} \right)+cos{3\varphi}\left(\begin{array}{c}
-s_5+u_9 \\ -t_5+v_9
\end{array} \right)
\right]\]
$$D=-3\pi\sin{4\varphi},$$
where, $\varphi=\pi/8$; the transfer matrix elements for calculation $x'$ and $y'$ at the characteristic reference points in Fig.~\ref{f4} via the input vector $(x
_1,x'_1)$ are used:
\[\left(\begin{array}{c}
x'_i \\ y'_i
\end{array} \right)=\left(\begin{array}{cc}
s_i & t_i \\
u_i & v_i
\end{array} \right)\cdot\left(\begin{array}{c}
x_1 \\ x'_1
\end{array} \right),~ ~ i=2,3,... 9;\]
In the equation for $D$ is taken into account the fact that the derivatives of the vertical ($\eta_y'$) and radial ($\eta_x'$) dispersion functions
all over the sections with solenoid insertions are zero in our case.
Note, the spin-orbit function $\vec d$ periodically depends on an azimuth. It always can be presented as a sum of two contributions: 
$$\vec d =\frac{\gamma d \vec{n}}{d \gamma}+\vec {d}_{beta}$$
where the first term is due to polarization axis chromaticity and the second one is due to betatron oscillations.
In our case the $\vec d$ function is expressed by way of the $A$, $B$ and $D$ coefficients. The terms in $\vec d$ including $A$ and $B$ present purely
the betatron contribution. Single, in a given case, term of $D$ origins from chromaticity of the $\vec n$ vector. 

\section{Estimate of radiative relaxation time}
\begin{figure}[htb]
\centering
\includegraphics*[width=120mm]{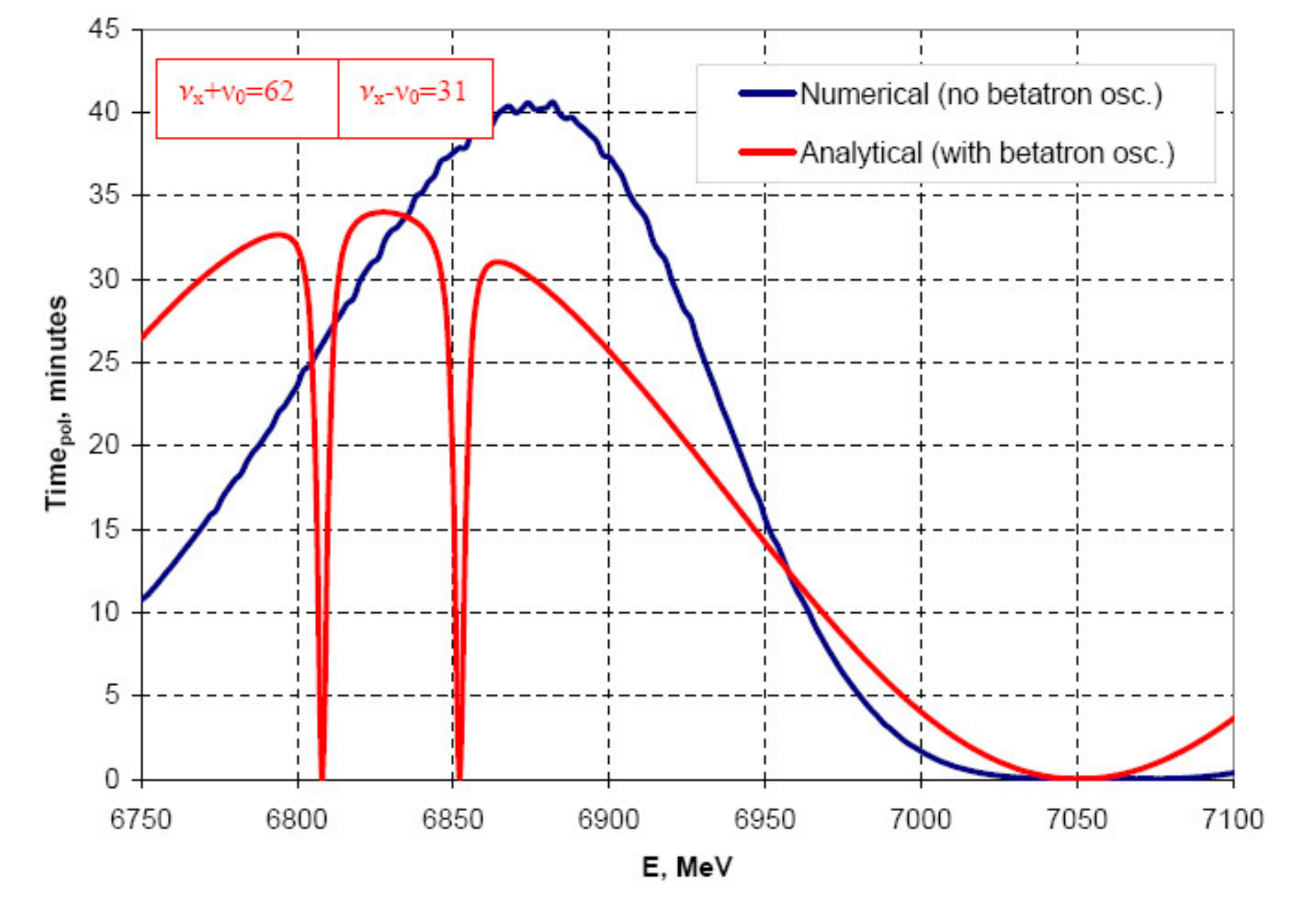}
\caption{\it Radiative relaxation rime of the polarization vs. the beam energy for the $2\times 3\times 5.7^0$ Central Arc Scheme.
Results of two approximations are presented. Types of depolarizing resonances are noticed. 
}
\label{f5}
\end{figure}

Energy dependence of the relaxation time has been found in two ways (see Fig.~\ref{f5}). 
In the first one (a blue line),  the polarization axis $\vec{n}$ in the main arcs as a function of azimuth and energy
is calculated using the spin matrix algebra similar to that at I.P. in Fig.~\ref{f1}.
Then numerical energy differentiation is performed to obtain a squared quantity of  a polarization vector chromaticity
averaged over main arcs: $<(\gamma d \vec{n}/d \gamma)^2>$.  Relaxation time is estimated from 
$$\tau_r\approx \frac{\tau_0}{1+\frac{11}{18}\left < \left (\frac{\gamma d \vec{n}}{d \gamma}\right )^2\right >}.$$

In the second way (a red line), with the aim to simplify a consideration of betatron contribution,  an approximation in which a spin rotation
angle over the Central Arc is assumed to be constant $=3\pi$ independently of an actual energy. Using the parameters of the solenoid insertion and FF region
optics the spin-orbit coupling coefficients A,B and D are calculated.  In our case
$$\beta_{x,1}=18.7m;$$
$$\alpha_{x,1}=0;$$
$$\eta_{x}=\eta'_x=0 ;$$
$$A=-6.7\times10^{-3}m^{-1};$$
$$B=-30;$$
$$D=-3\pi.$$
Estimation of $\tau_r$ using the coefficients obtained can be found in the form 
$$\tau_r\approx \frac{\tau_0}{1+\frac{11}{18}< \vec {d}^2>},$$
$$<\vec{d}^2>\approx \frac{<|h|^2> [(A\beta_{x,1})^2+B^2][1-\cos{2\pi\nu_0}\cos{2\pi\nu_x}]}{4 \beta_{x,1}\sin^2{\pi(\nu_x+\nu_0)} \sin^2{\pi(\nu_x-\nu_0)}}
+\frac{D^2}{4 \sin^2{\pi\nu_0}.} \eqno{(6)}$$
The value $<|h|^2>$ is found with typical values of the $\beta_x\approx5$ m and $\eta_x\approx 0.1$ m in the main arcs. 
Corresponding energy dependence of relaxation time is shown in Fig.5 by the red line.
 
It is seen from Fig.~\ref{f5} that maximums of the "blue" and "red" dependences do not coincide in position at energy. 
Because of the
approximation mentioned above, the dependence including betatron contribution has a quasi-symmetrical view relative to the characteristic
points 7050 and 6610 MeV
where the $\nu$ parameter becomes  integer. 
Approach using a numerical calculation of
$\gamma d \vec{n}/d \gamma$ reaches a maximal value of 40 minuts at 6870 MeV - nearer to the region with the nominal energy of 6958 MeV
Obviously, with closing to the characteristic points, the latter approach gives more accurate results. 
Nevertheless, the approach using the $A$, $B$ coefficients allows to estimate a reduction of maximum
of the interesting dependence as well as to determine the bandwidth of combination spin resonances ($\nu_0\pm\nu_x=k$).
The reduction can make in the point of maximum about 7 minuts. The combination resonance bandwidth is about 10 MeV. 

\section{Discussion}
The radiative relaxation time in the given scheme can achieve 30-35 minuts in the energy range of about 100 MeV near the nominal energy.
This corresponds to the average polarization degree $\bar{P}\approx 79 \%$ at $\tau_{\it l}=5$ minuts and $P_0=90 \%$. Dip of the polarization 
can be in the 10 MeV regions of the spin resonances $\nu_0\pm\nu_x=k$.  Such a result may be considered  as wholly satisfactory if omitting
additional depolarizing factors (spin-orbit coupling at the section with the Central Arc bend magnets,  magnetic field imperfections
and beam-beam effects in spin motion). Taking into account these factors one can conclude that
a margin of relaxation time is preferable.
For comparison, the alternative variant for SuperB \cite{bgmnik}
can provide the radiative relaxation time up to 2.7 hours near 7 GeV due to more optimal arrangement of the original solenoid insertions
in the same general magnetic structure of HER.  Under this arrangement, the Central Arc bend is 3 times smaller . This means decrease by a factor of 3
of the polarization axis chromaticity into the spin-orbit coupling coefficient $D$. 
At the same time, the chromatic effects concerning the beta-function bandwidth at I.P. are not minimized in the case
of the alternative variant. The latter circumstance does not allow to consider it accomplished.
 
\section{Summary}
It has been proved the Central Arc scheme for obtaining longitudial polarization at the 7 GeV SuperB HER can be considered as acceptable even in the variant
of large bend angle ($2\times 3\times 5.7 ^0$). More accurate next studies and development of the similar schemes with the aim to have a significant margin in the
radiative relaxation time of polarization as well as to fit the optic parameters in the best way are desirable. 

\section{Acknowledgements}
Author thanks Marchello Giorgi for showing interest in this work, Pantaleo Raimondi and Mikhail Zobov for discussions, Marica Biagini for assistance in the optics calculation.

\end{document}